\documentclass[%
 preprint,
 superscriptaddress,
 showpacs,
 amsmath,amssymb,
 aps,
 prl,
]{revtex4-1}

\usepackage{graphicx}% Include figure files
\usepackage{dcolumn}% Align table columns on decimal point
\usepackage{bm}% bold math

\begin{document}

\title{Structural evolution in the neutron-rich nuclei $^{106}$Zr and $^{108}$Zr 
}

\author{T.~Sumikama}
\email{sumikama@ph.noda.tus.ac.jp}

\author{K.~Yoshinaga}
\affiliation{Department of Physics, Tokyo University of Science, 2641 Yamazaki, 
Noda, Chiba 278-8510, Japan}

\author{H.~Watanabe}
\author{S.~Nishimura}
\affiliation{RIKEN Nishina Center, 2-1 Hirosawa, Wako, Saitama 351-0198, Japan}

\author{Y.~Miyashita}
\affiliation{Department of Physics, Tokyo University of Science, 2641 Yamazaki, 
Noda, Chiba 278-8510, Japan}

\author{K.~Yamaguchi}
\affiliation{Department of Physics, Osaka University, 1-1 
Machikaneyama,Toyonaka, Osaka 560-0043  Japan}

\author{K.~Sugimoto}
\author{J.~Chiba}
\affiliation{Department of Physics, Tokyo University of Science, 2641 Yamazaki, 
Noda, Chiba 278-8510, Japan}

\author{Z.~Li}
\affiliation{RIKEN Nishina Center, 2-1 Hirosawa, Wako, Saitama 351-0198, Japan}

\author{H.~Baba}
\affiliation{RIKEN Nishina Center, 2-1 Hirosawa, Wako, Saitama 351-0198, Japan}

\author{J.~S.~Berryman}
\affiliation{Nuclear Science Division, Lawrence Berkeley National Laboratory, 
Berkeley, CA 94720, USA}
\affiliation{NSCL, Michigan State University, East Lansing, Michigan 48824, USA}

\author{N.~Blasi}
\affiliation{INFN, Sezione di Milano, via Celoria 16, I-20133 Milano, Italy}

\author{A.~Bracco}
\author{F.~Camera}
\affiliation{INFN, Sezione di Milano, via Celoria 16, I-20133 Milano, Italy}
\affiliation{Dipartimento di Fisica, Universit\`a di Milano, via Celoria 16, I-
20133 Milano, Italy}

\author{P.~Doornenbal}
\affiliation{RIKEN Nishina Center, 2-1 Hirosawa, Wako, Saitama 351-0198, Japan}

\author{S.~Go}
\author{T.~Hashimoto}
\author{S.~Hayakawa}
\affiliation{Center for Nuclear Study (CNS), University of Tokyo, RIKEN campus, 
Saitama 351-0198, Japan}

\author{C.~Hinke}
\affiliation{Physik Department, Technische Universit\"{a}t M\"{u}nchen, D-85748 
Garching, Germany}

\author{E.~Ideguchi}
\affiliation{Center for Nuclear Study (CNS), University of Tokyo, RIKEN campus, 
Saitama 351-0198, Japan}

\author{T.~Isobe}
\affiliation{RIKEN Nishina Center, 2-1 Hirosawa, Wako, Saitama 351-0198, Japan}

\author{Y.~Ito}
\affiliation{Department of Physics, Osaka University, Machikaneyama 1-
1,Toyonaka, Osaka 560-0043  Japan}

\author{D.~G.~Jenkins}
\affiliation{Department of Physics, University of York, Heslington, York YO10 
5DD, United Kigdom}

\author{Y.~Kawada}
\author{N.~Kobayashi}
\author{Y.~Kondo}
\affiliation{Department of Physics, Tokyo Institute of Technology, 
O-Okayama, Meguro, Tokyo 152-8551, Japan}

\author{R.~Kr\"ucken}
\affiliation{Physik Department, Technische Universit\"{a}t M\"{u}nchen, D-85748 
Garching, Germany}

\author{S.~Kubono}
\affiliation{Center for Nuclear Study (CNS), University of Tokyo, RIKEN campus, 
Saitama 351-0198, Japan}

\author{G.~Lorusso}
\affiliation{RIKEN Nishina Center, 2-1 Hirosawa, Wako, Saitama 351-0198, Japan}
\affiliation{NSCL, Michigan State University, East Lansing, Michigan 48824, USA}

\author{T.~Nakano}
\affiliation{Department of Physics, Tokyo University of Science, 2641 Yamazaki, 
Noda, Chiba 278-8510, Japan}

\author{M.~Kurata-Nishimura}
\affiliation{RIKEN Nishina Center, 2-1 Hirosawa, Wako, Saitama 351-0198, Japan}

\author{A.~Odahara}
\affiliation{Department of Physics, Osaka University, Machikaneyama 1-
1,Toyonaka, Osaka 560-0043  Japan}

\author{H.~J.~Ong}
\affiliation{Research Center for Nuclear Physics (RCNP), Osaka University, 
Ibaraki, Osaka 567-0047, Japan}

\author{S.~Ota}
\affiliation{Center for Nuclear Study (CNS), University of Tokyo, RIKEN campus, 
Saitama 351-0198, Japan}

\author{Zs.~Podoly\'ak}
\affiliation{Department of Physics, University of Surrey, Guildford GU2 7XH, 
United Kingdom}

\author{H.~Sakurai}
\affiliation{RIKEN Nishina Center, 2-1 Hirosawa, Wako, Saitama 351-0198, Japan}

\author{H.~Scheit }
\affiliation{RIKEN Nishina Center, 2-1 Hirosawa, Wako, Saitama 351-0198, Japan}

\author{K.~Steiger}
\affiliation{Physik Department, Technische Universit\"{a}t M\"{u}nchen, D-85748 
Garching, Germany}

\author{D.~Steppenbeck }
\affiliation{RIKEN Nishina Center, 2-1 Hirosawa, Wako, Saitama 351-0198, Japan}

\author{S.~Takano}
\affiliation{Department of Physics, Tokyo University of Science, 2641 Yamazaki, 
Noda, Chiba 278-8510, Japan}

\author{A.~Takashima}
\author{K.~Tajiri}
\affiliation{Department of Physics, Osaka University, Machikaneyama 1-
1,Toyonaka, Osaka 560-0043  Japan}

\author{T.~Teranishi}
\affiliation{Department of Physics, Kyushu University, Fukuoka 812-8581, Japan}

\author{Y.~Wakabayashi}
\affiliation{Japan Atomic Energy Agency, Tokai, Ibaraki 319-1195, Japan}

\author{P.~M.~Walker}
\affiliation{Department of Physics, University of Surrey, Guildford GU2 7XH, 
United Kingdom}

\author{O.~Wieland}
\affiliation{INFN, Sezione di Milano, via Celoria 16, I-20133 Milano, Italy}

\author{H.~Yamaguchi}
\affiliation{Center for Nuclear Study (CNS), University of Tokyo, RIKEN campus, 
Saitama 351-0198, Japan}

\date{April 12, 2011}

\begin{abstract}
The low-lying states in $^{106}$Zr and $^{108}$Zr have been investigated 
by means of $\beta$-$\gamma$ and isomer spectroscopy at the RI beam factory, respectively.
A new isomer with a half-life of $620\pm150$ ns has been identified in $^{108}$Zr. 
For the sequence of even-even Zr isotopes, the excitation energies of 
the first $2^+$ states reach a minimum at $N=64$ and gradually increase 
as the neutron number increases up to $N=68$,   
suggesting a deformed sub-shell closure at $N=64$.
The deformed ground state of $^{108}$Zr indicates 
that a spherical sub-shell gap predicted at $N=70$ is not large
enough to change the ground state of $^{108}$Zr to the spherical shape. 
The possibility of a tetrahedral shape isomer in $^{108}$Zr is also discussed. 
\end{abstract}


\maketitle

%%%%%%%%%%%%%%%%%%%%%%%%%%%%%%%%%%%%%%%%%%%%%%%%%%%%%%%%%%%%%%%%%%%%%%
Atomic nuclei, which are strongly-interacting finite quantum systems, 
 have a spherical or deformed shape.  
The shape evolution as a function of proton and neutron numbers  
is related to the shell structure for spherical or deformed shape.  
At proton number $Z=40$, the shell structures for a wide variety of shapes, i.e.~spherical, prolate, oblate, and more exotic tetrahedral shapes, are closed \cite{Skalski97, Schunck04}.  
So, shape transitions and maximum deformation at deformed sub-shell closure are expected  for Zr isotopes ($Z=40$). 
%The shell structure for nuclei far from the stability line can be investigated through the shape evolution of Zr isotopes. 

The shape of  neutron-rich Zr isotopes changes drastically 
from spherical to  deformed at neutron number $N=60$ \cite{Cheifetz70}. 
The quadrupole deformation is known to increase toward $N=64$ 
from half-life measurements of the first $J^{\pi}=2^+$ ($2_1^+$) state of even-even Zr isotopes \cite{Raman01, Hwang06}. 
However, the  evolution of the deformation beyond $N=64$ is unknown 
because there is no spectroscopic information. 
Several authors have predicted 
different types of shape transition in the more neutron-rich region around $N=70$. 
Specifically, a ground-state shape transition from prolate to oblate is 
predicted at $N=72$ \cite{Xu02} or at  $N=74$ \cite{Skalski97}. 
In contrast, a spherical sub-shell gap at $N=70$ and 
a spherical shape of $^{110}$Zr are predicted 
by the Skyrme energy density functional with a tensor 
force and a reduced spin-orbit term \cite{Bender09}. 
A spherical shape might be expected to appear also in Zr isotopes approaching $^{110}$Zr. 
Furthermore, an exotic shape with tetrahedral symmetry is 
predicted to be stabilized around $^{110}$Zr, 
which has the tetrahedral magic numbers, $Z=40$ and $N=70$   \cite{Schunck04}. 
An excited state with the tetrahedral shape, predicted at $^{108}$Zr, 
may become an isomeric state. 
The tetrahedral shape is still hypothetical in atomic nuclei, 
in spite of many recent theoretical and experimental works \cite{Jentschel10}. 

In addition to these striking shape effects, very neutron-rich Zr isotopes are important for understanding nucleosynthesis 
through the rapid neutron-capture process (r process). 
The estimated r-process abundances between $A=110$ - 125 depend
on nuclear mass models \cite{Wanajo06}.
Here, the astrophysical model of prompt supernova explosions with O-Ne-Mg cores is applied. 
The abundance deficiencies by one order of magnitude have been reduced 
when mass models including shell quenching at $N = 82$ are used.
A more drastic shell evolution, accompanied by not only the $N=82$ shell quenching but also the new $N=70$ sub-shell, may influence the r-process path as predicted by self-consistent mean-field calculations \cite{Dobaczewski94, Bender09}. 
Thus, in order to predict nuclear properties reliably for $A\approx 110$ - 125 
on the r-process path (i.e.~$N=70$ - 82), 
it is crucial to know whether the sub-shell gap at $N=70$ is large or not.  
The $N=70$ sub-shell effect can be investigated through the structural evolution of Zr isotopes near $N=70$. 

$\beta$-decay half-lives can be used as a probe of shape transitions even though their values depend mainly on the $\beta$-decay $Q$ values. For example,
calculations using the quasiparticle random-phase approximation (QRPA) 
predict shorter $\beta$-decay half-lives for a spherical shape than for a deformed one \cite{Sarriguren10}. 
The half-lives measured previously, including $^{110}$Zr \cite{Pereira09, Nishimura11}, however, show a gradual decrease as a function of the neutron number, with no indication of the transition to spherical shape. 

The purpose of this Letter is to study the structural evolution caused by spherical and deformed shell structures in the neutron-rich Zr isotopes
through the low-lying states obtained from $\beta$-$\gamma$ and isomer spectroscopy of $^{106}$Zr and $^{108}$Zr, 
and to discuss a new isomeric state of $^{108}$Zr in the context of the possibility of a tetrahedral shape isomer. 
The $2^+_1$ state energy, $E(2_1^+)$, and 
the ratio of the first 4$^+$ to 2$^+$ state energies, $R_{4/2} = E(4_1^+)/E(2_1^+)$, 
for even-even nuclei serve as valuable indicators of 
deformation and shell closure \cite{Cakirli08}.

The experimental studies of $^{106}$Zr and $^{108}$Zr were performed at RIBF 
operated by the RIKEN Nishina Center and the CNS of the University of Tokyo. 
Secondary beams were produced using in-flight fission of $^{238}$U beams having 
an energy of 345 MeV/nucleon. 
The Be production target  was 3 mm thick. 
Fission fragments were separated using the RI-beam separator, BigRIPS 
\cite{Kubo03, Ohnishi08}. 
For isotopic separation, an aluminum wedge-shaped energy degrader having a 
median thickness of 5.8 mm was placed 
at the momentum dispersive focus of the first stage of BigRIPS. 
An additional degrader was placed 
at the momentum dispersive focus of the second stage of BigRIPS
in order to eliminate the fragments 
that were not fully stripped before reaching the second degrader. 
The shape of the degrader was designed to satisfy the momentum achromatic 
condition at the next achromatic focus. 
The thickness along the center of the beam line was 2.1 mm. 
The separated particles were transported through 
the ZeroDegree spectrometer \cite{Ohnishi08} 
to the final focal plane of ZeroDegree. 

Beam particles were identified using the magnetic rigidity, $B\rho$, 
time-of-flight, TOF, 
and energy loss, $\Delta E$, determined using the focal plane detectors of 
BigRIPS and ZeroDegree  \cite{Watanabe11}. 
The particle identification spectrum for the same data set is shown in Ref.~\cite{Watanabe11}. 
The resolution of the atomic number $Z$ and the mass-to-charge ratio $A/Q$ for $^{108}$Zr was 
0.4 and 0.005 (full width at half maximum), respectively.  
The identified particles were implanted into 
nine stacked double-sided silicon strip detectors (DSSD). 
Each DSSD had a thickness of 1 mm with an active area of $50 \times 50$ mm$^2$ 
segmented into $16 \times 16$ strips. 
The implanted particles and subsequent $\beta$ rays were detected by the DSSDs. 
The DSSDs were surrounded by four Compton-suppressed Clover-type Ge detectors 
and two LaBr$_3$ detectors, 
which were used to detect $\gamma$ rays following $\beta$ and isomeric decays. 
A plastic scintillation detector was placed in front of 
each Clover detector. 
By taking an anti-coincidence between the plastic and Clover detector, 
the background arising from $\beta$-ray events in the $\gamma$-ray spectrum
can be eliminated.

\begin{figure}
\includegraphics[width = 82 mm]{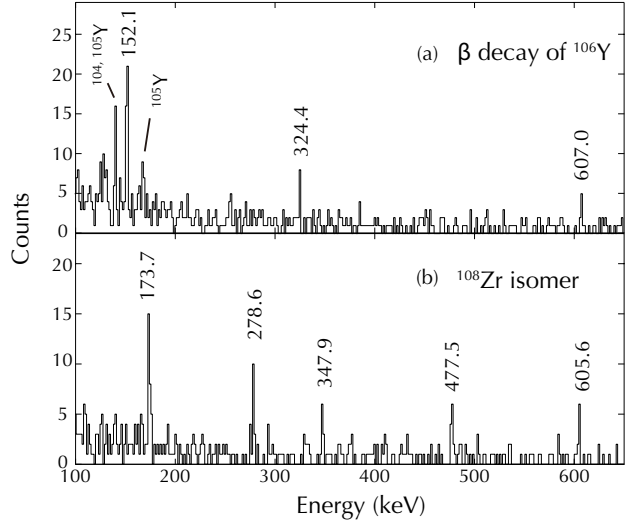}
\caption{
Gamma-ray spectra measured (a) in coincidence with $\beta$ rays detected 
within 200 ms after implantation of $^{106}$Y and (b) with a particle gate on 
$^{108}$Zr within 4 $\mu$s.
Peaks marked with the nucleus name indicate ones measured also in coincidence 
with the $\beta$ decay of its nucleus. 
}  
\label{106Zr}\label{108Zr}
\end{figure}
\begin{figure}
\includegraphics{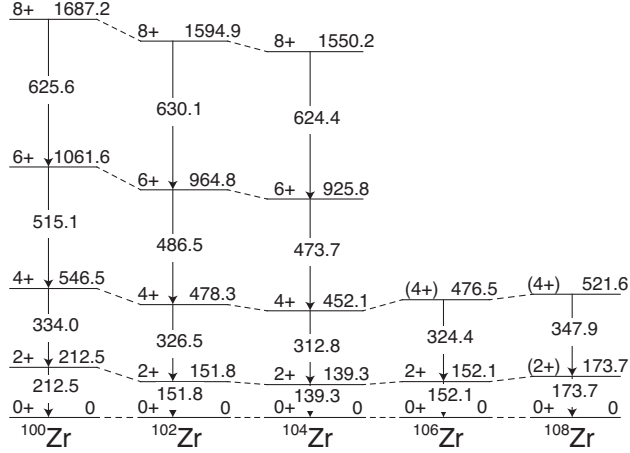}
\caption{Ground-state bands of neutron-rich even-even Zr isotopes with $N \ge 
60$. 
The energies of $^{100 - 104}$Zr are taken from the ENSDF database \cite{ENSDF}. 
 }
\label{ZrLevels}
\end{figure}

$\beta$-decay events were selected 
using the position and time correlations between the implanted particle and the 
$\beta$ ray. 
The relative position was restricted to the same pixel of the DSSD.
The $\beta$-decay half-life of $^{106}$Y was measured to be $62^{+25}_{-14}$ ms \cite{Nishimura11} 
in the same data set. 
Figure ~\ref{106Zr} (a) shows  a $\gamma$-ray spectrum 
measured in coincidence with $\beta$ rays following the implantation of $^{106}$Y
within 200 ms, which is about three times longer than the $\beta$-decay half-life. 
The $\gamma$ ray with 140 keV was also measured in coincidence with the $\beta$ 
decay of $^{104}$Y and $^{105}$Y, which were included in the cocktail beam 
\cite{Watanabe11}. 
This $\gamma$ ray was emitted from the $2^+_1$ state in $^{104}$Zr  
after $\beta$-delayed two-neutron emission of $^{106}$Y. 
The $\gamma$ ray with 169 keV was also measured in coincidence with the $\beta$ 
decay of $^{105}$Y  
and results from $^{105}$Zr after $\beta$-delayed one-neutron emission.  
The most intense $\gamma$ ray at 152 keV was assigned as the transition 
from the $2_1^+$ state to the ground state in $^{106}$Zr. 
Figure \ref{ZrLevels} shows the ground-state bands of the even-even Zr isotopes 
with $N \ge 60$. 
The gradual evolution of $E(2_1^+)$ supports the $^{106}$Zr assignment. 

The spin and parity of the parent nucleus $^{106}_{\ 39}$Y$_{67}$ are possibly 
$2^+$ or $3^+$, 
because the ground states of $^{99, 101}$Y 
are indicated to have the same proton configuration, $5/2^+$[422], as  
$^{101, 103, 105}$Nb \cite{Hotchkis91,Cheal07,Cheal09} 
and the spin and parity of $^{108}_{\ 41}$Nb$_{67}$ is suggested to be $2^+$ or $3^+$ 
\cite{Penttila96}.
The $4^+_1$ and the second $2^+$ ($2^+_2$) states of $^{106}$Zr 
are likely to be populated in the $\beta$ decay of $^{106}$Y 
by comparison with the population of the $4^+_1$ and $2^+_2$ states of $^{108}$Mo
in the $\beta$ decay of $^{108}$Nb \cite{Penttila96}. 
If the 324 keV $\gamma$ ray is the transition to the $2^+_1$ state, 
the excited state energy is 477 keV. 
Since $E(2_1^+)$ of $^{106}$Zr is slightly larger than that of $^{104}$Zr (Fig.~\ref{ZrLevels}), 
$E(4_1^+)$ is expected to increase gradually and to be 450 - 500 keV. 
The energies of the $4^+_1$ and $2^+_2$ states of $^{106}$Zr are predicted  
by using the interacting boson model (IBM) \cite{Lalkovski09}.  
The parameters of the IBM are obtained from a least-squares fit to the known level energies
of $^{108}$Mo, $^{110}$Ru, and $^{112}$Pd along the isotonic chain ($N=66$). 
The largest deviations between the experimental and theoretical $E(4_1^+)$ and $E(2_2^+)$ states
are 34 keV and 76 keV, respectively. 
The $E(4_1^+)$ and $E(2_2^+)$ of $^{106}$Zr are extrapolated 
to be 455 keV and 618 keV, respectively. 
Therefore, the excited states at 477 keV and 607 keV were tentatively assigned 
as the $4_1^+$ and $2_2^+$ states in $^{106}$Zr, respectively. 
The transition from the $2_2^+$ state to the $2_1^+$ state is expected,  
but no $\gamma$-ray peak at 455 keV was observed due to the low statistics.  

The $\gamma$ rays emitted from a new isomeric state of $^{108}$Zr 
were observed within 4 $\mu$s after the implantation  of $^{108}$Zr
as shown in Fig.~\ref{108Zr} (b). 
Five $\gamma$-ray peaks at energies of 174, 279, 348, 478, and 606 keV 
were unambiguously measured. 
A half-life of $620 \pm 150$ ns was derived from the sum of time spectra for 
these five $\gamma$ rays.  
Some low-intensity $\gamma$-ray peaks from the $^{108}$Zr isomer might not 
have been identified, and no information on $\gamma$-$\gamma$ coincidences 
was obtained due to the low statistics. Nevertheless, it can be estimated that
the energy of the isomeric state is likely more than 1 MeV.  
The ground-state band is populated up to $4^+$, thus the spin is likely more than or equal to 4. 
Before discussing possible structures of the observed isomer, 
low-lying states of $^{108}$Zr are discussed.  

If a spherical ground state would appear around $^{110}$Zr 
due to the predicted $N=70$ sub-shell gap \cite{Bender09},   
then $E(2_1^+)$ would have to suddenly increase and $R_{4/2}$ drop to $\approx 2$. 
However, $E(2_1^+)$ of $^{106}$Zr is similar to that of Zr isotopes with $A=100 
- 104$,  which are well deformed with $\beta_2 =  0.355(10)$, $0.43(4)$,  and 0.47(7) 
for $A=100$, $102$, and 104, respectively \cite{Raman01,Hwang06}.   
Because the $\gamma$-ray energies of 174 and 348 keV in $^{108}$Zr
are slightly larger than those of 152 and 324 keV in $^{106}$Zr
and the relevant energies smoothly change from $^{100}$Zr to $^{108}$Zr (Fig.~\ref{ZrLevels}), 
the 174 and 348 keV $\gamma$ rays were tentatively assigned as
the transitions from the $2^+_1$ state to the ground state and 
from the $4^+_1$ state to the $2^+_1$ state, respectively.   
$R_{4/2}$ gradually changes with values of 2.57, 3.15, 3.25, 3.13, and 3.00 
for $A=100, 102, 104, 106,$ and 108, respectively. 
Values of $R_{4/2}$,  which is close to 3.3 for a rigid rotor,  
indicate the rotational character of a deformed nucleus.   
The ground state of $^{108}$Zr is most likely as deformed as $^{106}$Zr.    
Therefore, the spherical sub-shell gap at $N=70$ seems not to be large enough
to change the ground state of $^{108}$Zr to spherical shape. 

\begin{figure}
\includegraphics{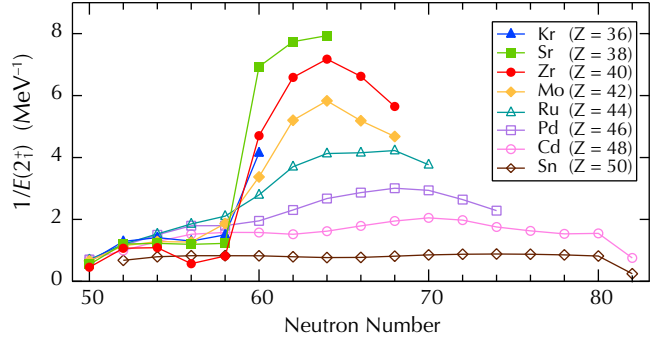}
\caption{(Color online) $1/E(2_1^+)$  as a function of the neutron number.
The present results are for Zr isotopes with $N=66$ and 68. 
Others are taken from the ENSDF database \cite{ENSDF}. }
\label{syst}
\end{figure}

The structural evolution around the neutron-rich Zr isotopes can be
visualized using $1/E(2^+_1)$ \cite{Cakirli08}.  
Figure~\ref{syst}  shows $1/E(2^+_1)$ as a function of the neutron number. 
The values of $1/E(2^+_1)$ suddenly increase at $N=60$ for Kr, Sr, Zr and Mo 
isotopes because of the onset of deformation. 
$1/E(2^+_1)$ reaches a maximum at $N=64$ for both Zr and Mo isotopes. 
Another remarkable behavior at $N=64$ has been  observed for Mo isotopes.
Hua \textit{et al}.\ observed a band crossing 
due to the rotation alignment of an $h_{11/2}$ neutron pair \cite{Hua04}. 
The shift of the band crossing to higher rotational frequency  in $^{106}$Mo
is interpreted as a consequence of
the deformed sub-shell closure at $N=64$. 
The maximum of $1/E(2^+_1)$ at $N=64$ can also be interpreted as being due to
the deformed sub-shell closure at $N=64$ with $\beta_2 \approx 0.47(7)$ \cite{Hwang06} 
for $^{104}$Zr. 

The r-process path between $A=110$ - 125 may be affected by 
the weakening of the spin-orbit force, which is associated with the neutron skin \cite{Pfeiffer96}. 
The harmonic-oscillator-like doubly magic nucleus of $^{110}$Zr \cite{Pfeiffer96} or the shell closure at $N=70$ \cite{Bender09}
is predicted by using an artificially reduced spin-orbit force. 
However, the shape of $^{108}$Zr as deformed as lighter Zr isotopes 
indicates no drastic evolution. 
One may conclude that the deficiencies of the estimated r-process 
abundances around $A=110$ are not caused by the drastic weakening of the spin-orbit force. 

The existence of a long-enough-lived ($T_{1/2} > 100$ ns) isomer gives access to 
the excited-state structure. 
The  $^{108}$Zr isomer is the only isomer discovered in even-even nuclei in the present work,  
although the production yield of the isotopes $^{104, 106}$Zr and the isotone $^{110}$Mo
was higher than $^{108}$Zr. 
It seems that the structure of the $^{108}$Zr isomer is suddenly stabilized.

A possible explanation of the isomerism is that the isomeric state of $^{108}$Zr has a tetrahedral shape.  
The tetrahedral shape is a non-axial octupole deformation 
coupled with a vanishingly small quadrupole deformation, 
therefore a $\gamma$ decay to normally quadrupole-deformed states is hindered  \cite{Jentschel10}. 
The tetrahedral shape will appear only when the competing shell effect for spherical shape is weak \cite{Dudek03}. 
This requirement might be satisfied because the $^{108}$Zr results indicate the deformed ground state.  
Furthermore, the energy barrier against different shapes plays an important role for the stability. 
The total energies of the tetrahedral or quadrupole deformed shape are calculated using the microscopic-macroscopic method with pairing correlations \cite{Schunck04}.  
The energy barrier between the tetrahedral and oblate shapes is predicted for $^{108}$Zr,  
but becomes very small for $^{104, 106}$Zr.  
The onset  of long-lived isomerism at $^{108}$Zr may be caused by the difference of the energy barrier.
The excitation energy $E_x$ is predicted to be 1.1 MeV by the Hartree-Fock-Bogoliubov (HFB) calculation \cite{Schunck04}. 
The spin $J$ of a possible band head is expected to be 3 \cite{Schunck04}, and 
the spin of the isomeric state might be larger than that of the band head \cite{Dudek03}. 
These expectations are consistent with the experimental indications, i.e.\ $E_x > 1$ MeV and $J \ge 4$. 
While the parity of the tetrahedral shape is predicted to be negative \cite{Zberecki09}, 
there is no experimental indication.

Another possibility of the isomeric state is a two-quasineutron state with high $K$ value, 
which is predicted 
for several even-even nuclei around $^{108}$Zr \cite{Xu02}.   
Considering the gradual change of $E(2_1^+)$ around $^{108}$Zr as shown in Fig.~\ref{syst}, 
this kind of isomeric state is expected to have a similar half-life and 
be populated not only in $^{108}$Zr 
but also in neighboring even-even nuclei.  
Thus, we propose the observed isomer is a promising candidate for the tetrahedral shape isomer rather than the high-K isomer. 

In summary, 
decay spectroscopy has been performed to assign the first $2^+$ and $4^+$ states 
of $^{106, 108}$Zr. 
The systematics of $E(2_1^+)$ indicate that 
the deformation reaches a  maximum at $N=64$ for Zr isotopes, 
suggesting  
a deformed sub-shell closure at $N=64$. 
In addition, the deformed ground state of $^{108}$Zr 
indicates that the spherical $N=70$ sub-shell gap is not having a large effect at $N=68$ for Zr isotopes. 
For a definite conclusion regarding the $N=70$ sub-shell, 
future measurements of $^{110}$Zr are required. 
A long-lived ($T_{1/2} > 100$ ns) isomer, possibly expected in even-even Zr nuclei, was discovered 
only in $^{108}$Zr. 
The isomeric state of $^{108}$Zr is proposed to be a candidate for a tetrahedral shape isomer  
because the energy barrier between the tetrahedral and oblate shapes is predicted to be more robust for $^{108}$Zr, compared with $^{104, 106}$Zr. 
To confirm whether this isomer has the tetrahedral shape or not, 
the determination of the decay scheme including spins and parities, and 
the measurement of the band structure above the isomer are required 
from future measurements. 

% Acknowledgment
The authors acknowledge the accelerator staff and the members of the BigRIPS 
team 
for their efforts in delivering the $^{238}$U and RI beams. 
This work was supported by KAKENHI (50126124 and 19340074),
RIKEN President Fund 2005, the DFG (EXC 153, KR 2326/2),  UK STFC and 
AWE plc. 

\providecommand{\noopsort}[1]{}\providecommand{\singleletter}[1]{#1}%

\end{document}